%% file: main.tex
\documentclass[letterpaper]{article}
\usepackage[utf8]{inputenc}
\usepackage{amsthm}
\usepackage{amssymb}
\usepackage{amsfonts}
\usepackage{amsmath}
\usepackage{calrsfs}
\usepackage{xcolor}
\usepackage{mathrsfs} 
\usepackage{nicefrac,xfrac}

\def\year{2021}\relax
\usepackage{aaai}  
\usepackage{times}  
\usepackage{helvet} 
\usepackage{courier}  
\usepackage[hyphens]{url}  
\usepackage{graphicx} 
\def\UrlFont{\rm}  
\usepackage{natbib}  
\usepackage{caption} 
\DeclareMathAlphabet{\pazocal}{OMS}{zplm}{m}{n}

\newcommand{\aj}{\mathbf{a}}

\newcommand{\A}{\pazocal{A}}
\newcommand{\x}{\mathbf{x}}

\newcommand{\z}{\mathbf{z}}

\newcommand{\R}{\mathbb{R}}

\newcommand{\eBNIC}{$\varepsilon$-BNIC}

\newcommand{\inter}{I^d \times I^s}

\newcommand{\caughtLying}{\pazocal{L}}
\newcommand{\unf}{\textsc{uniform}}

\newcommand{\unfk}{\textsc{uniform-k}}

\renewcommand{\cite}{\citep}

\newtheorem{theorem}{Theorem}[]
\newtheorem{definition}{Definition}[]
\newtheorem{corollary}{Corollary}[]
\newtheorem{remark}{Remark}[]
\newtheorem{example}{Example}[]
\newtheorem{proposition}{Proposition}

\begin{document}
\title{Incentivizing Truthfulness Through Audits in Strategic Classification}
\author{Andrew Estornell\\Computer Science \& Engineering\\Washington University in St.~Louis\\aestornell@wustl.edu\\ 
\And Sanmay Das\\Computer Science\\George Mason University\\sanmay@gmu.edu \And Yevgeniy Vorobeychik\\Computer Science \& Engineering\\Washington University in St.~Louis\\yvorobeychik@wustl.edu}

\maketitle

\begin{abstract}
    In many societal resource allocation domains, machine learning methods are increasingly used to either score or rank agents in order to decide which ones should receive either resources (e.g., homeless services) or scrutiny (e.g., child welfare investigations) from social services agencies. An agency's scoring function typically operates on a feature vector that contains a combination of self-reported features and information available to the agency about individuals or households.
    This can create incentives for agents to misrepresent their self-reported features in order to receive resources or avoid scrutiny, but agencies may be able to selectively audit agents to verify the veracity of their reports.
    
    We study the problem of optimal auditing of agents in such settings.
    When decisions are made using a threshold on an agent's score, the optimal audit policy has a surprisingly simple structure, uniformly auditing all agents who could benefit from lying.
    While this policy can, in general be hard to compute because of the difficulty of identifying the set of agents who could benefit from lying given a complete set of reported types, we also present necessary and sufficient conditions under which it is tractable.
    We show that the scarce resource setting is more difficult, and exhibit an approximately optimal audit policy in this case.
    In addition, we show that in either setting verifying whether it is possible to incentivize exact truthfulness is hard even to approximate.
    However, we also exhibit sufficient conditions for solving this problem optimally, and for obtaining good approximations.
  
\end{abstract}

\input{Introduction.tex}



\input{Model.tex}

\input{NeedForAudit.tex}

\input{Optimal_Audit_Policy.tex}

\input{Hardness.tex}




\input{Special_Cases.tex}

\input{conclusion.tex}
\subsection*{Acknowledgments}

This research was partially supported by the National Science Foundation (IIS-1910392, IIS-1939677, IIS-1905558, IIS-1903207, IIS-1927422, and ECCS-2020289), Army Research Office (W911NF-19-1-0241), and Amazon.

\newpage
\bibliographystyle{aaai21}
\bibliography{truth}
\newpage

\end{document}

%% file: Introduction.tex
\section{Introduction}

Algorithmic decision-making systems are increasingly used to make high-stakes resource allocation decisions by social services agencies. This includes both scarce resource settings, where the demand for a limited pool of resources exceeds supply (for example, housing for the homeless \cite{kube2019allocating}), as well as risk-scoring settings, where only those who fall above or below a certain threshold are either given a resource (for example, a loan \cite{agarwal2009payday}) or targeted for further scrutiny (for example, parents suspected of child maltreatment or neglect \cite{chouldechova2018case}). 
As is standard in classification and ranking settings, each individual or household (henceforth \emph{agent}) is associated with a feature vector. In many such settings, the feature vector will combine information submitted by the agents themselves with information about them available from other sources. For example, in prioritizing households for homeless services, agencies make decisions based on self-reported items (e.g., history of alcohol or drug use) as well as on information available to them in government records (e.g., child-support or welfare payments received) \cite{brown2018reliability}. Naturally, this creates incentives for agents to try and game the system by strategically choosing their self-reported features in order to maximize their chances of receiving the resource or avoiding scrutiny. 

Prior work on \emph{strategic} or \emph{adversarial classification} has considered a closely related problem where agents subject to classification can modify feature values at some cost or subject to a constraint on the total magnitude of such modification, with the goal of inducing an incorrect prediction~\cite{Athalye18,Carlini17,Hardt15,Milli19,Papernot18,Tong19,Vorobeychik18book}.
This research has typically focused on either assessing how vulnerable particular families of classifiers are to such attacks (often termed \emph{adversarial examples})~\cite{Athalye18,Carlini17,kdd2005,ndss2016}, or on designing classifiers that are robust in the sense that the prediction remains unchanged even after budget-constrained feature modifications~\cite{kdd2011,Bruckner12,Hardt15,li2016,Madry18,Tong19,Wong18}.
In this literature, the interests of the agents are commonly viewed as opposed to those of the decision-maker (e.g., learner), often motivated by security considerations~\cite{oakland2014,ndss2016}.
Moreover, the typical models representing costs to agents of modifying features are at times not adequate at capturing realistic limits on what agents can do~\cite{Tong19,Wu20}.
In contrast, in the kinds of social services settings we describe, and potentially numerous others (e.g., tax filing), the costs of misrepresenting one's self-reported features are better captured by the risk associated with being audited than, say, a hard constraint on how much the features are modified.
Moreover, the agents' interests are not fundamentally opposed to the principal's; rather, this is a case of misaligned incentives more akin to that studied in the incentive design literature~\cite{Haeringer18,Nisan07}.

We consider a principal who has a limited budget of audits and can use these to determine whether an agent is telling the truth, with the cost of failing an audit the primary deleterious consequence to dishonest agents.
For example, caseworkers can interview associates of the agent and ask about behavioral issues, alcohol or drug use, and the like, and impose restrictions or fines on the agent if the results reveal dishonesty.
We suppose that the principal uses a score function $f$ (for example, learned risk scores) that takes agent features as input in order to decide whether an agent is subject to further scrutiny whenever their score exceeds a predefined threshold (we term this the \emph{threshold} setting), or to allocate resources to the agents with the top $k$ values of $f$ (we call this the \emph{top-$k$} setting).
We specifically focus on two problems: 1) designing an audit policy for the principal that minimizes incentives to lie, defined in terms of approximate Bayes-Nash incentive compatibility ($\varepsilon$-BNIC), and 2) verifying whether it is possible to ensure truthful reporting of features.
We show that in the threshold allocation setting an optimal policy audits uniformly at random all agents who are above the threshold, with special consideration for those who are either obviously lying or telling the truth.
Although this policy is in general hard to compute, we present sufficient conditions under which it is tractable.
In the top-$k$ setting, we prove that auditing all agents who receive the scarce resource uniformly at random (again, modulo special treatment of agents who are either certainly truthful or dishonest) yields an additive approximation bound, although the problem is hard in general.
Furthermore, we show that this audit policy is optimal if we consider dominant strategy incentive compatibility as a solution concept instead of \eBNIC.

Surprisingly, the verification problem is even harder: determining if any audit policy can incentivize truthful reporting is \#P-hard even for a uniform prior over features and only two agents.
However, we give sufficient conditions under which verification becomes tractable in the threshold setting for both piecewise linear and logistic scoring functions.
Our corresponding results are weaker for the top-$k$ setting, where we require the distribution over features to be uniform to obtain a tractable algorithm for checking incentives to lie \emph{assuming} that a uniform audit policy is used.
Finally, we show that for distributions for which we can efficiently approximate integrals over intervals, we can also approximately verify incentive compatibility.

Our results are important for understanding the potential for audits to be useful in various social services settings. Of perhaps the most practical importance is the clear distinction we find between the threshold (modeling unlimited, but costly, deployment of resources) and top-k (modeling scarce resource allocation) settings in terms of the difficulty of finding a good audit policy, and the simplicity of the optimal audit policy in the threshold setting.



%% file: Model.tex
\section{Preliminaries}

We consider a setting with a collection of $n$ agents in which either a scarce resource is distributed among $k$ of them using a score function, or each agent is scored to determine whether they are selected to receive a resource.
Each agent is associated with a vector of attributes (features) which are grouped into two categories: ``known'', denoted by $\x$, and ``self-reported'', denoted by $\z$.
Throughout, we refer to $(\x,\z)$ as an agent's \emph{true type}, to contrast it with $(\x,\z')$ in which $\z'$ is self-reported and may be different from the true corresponding characteristics of the agent.
For example, the agent may have a history of substance abuse, corresponding to ``true" $z_j = 1$, but reports that they do not, with ``reported" $z_j' = 0$.
Let $d$ be the number of known and $s$ the number of self-reported features.
We assume that each feature in either category either belongs to a continuous or discrete interval, i.e., each $x_j, z_k \in I = [a,b] \cap S$, where $S = \R$ (continuous interval) or $S = \mathbb{Z}$ (discrete interval).
We further assume that the true types of each of the $n$ agents are i.i.d.\ according to a (common knowledge) prior distribution $D$ with PDF (or PMF, in the discrete case) denoted by $h:I^d\times I^s \rightarrow [0, 1]$.
We will use $\mathbb{P}(\cdot)$ to denote the associated probability measure.




Let $\pazocal{A} = \{\aj_1, ..., \aj_n\}$ denote the collection of $n$ agents, where $\aj_i = (\x_i, \z_i)\in I^d\times I^s$ represents the agent's \emph{true} type, and let $\pazocal{A}' = \{\aj_1', ..., \aj_n'\}$ be the collection of reported types, $\aj_i' = (\x_i, \z_i')$. We assume that each agent knows their own type, but only knows the common prior $h$ about the types of other agents.

The principal publishes a score function $f:I^d \times I^s \rightarrow \R$ that takes each agent's \emph{reported} type $\aj_i'$ as input, and returns a real-valued score.
For example, $f$ may represent the probability (learned from historical data) that a homeless person will be safely and stably housed in 1 year if allocated a housing resource.
There are two common ways that $f$ is used in resource allocation: (1) \textbf{Threshold allocation: } all agents scoring above a threshold $\theta$ are allocated a resource (e.g., not chosen for further scrutiny in a child neglect case), and (2) \textbf{Top-$k$ allocation: } agents with the highest $k$ scores based on reported types are allocated a resource (e.g., housing). 

The principal can \emph{audit} up to $B$ agents and thereby verify whether their reported type matches their true type.
Let $\phi$ denote the audit policy, which is a function of the full collection of $n$ reported types $\pazocal{A}'$.
We consider stochastic audit policies, where $\phi_i(\pazocal{A}') \in [0, 1]$ is the probability that agent $i$ is audited.
If an audit of agent $i$ determines that the agent has lied, i.e., $\z_i' \ne \z_i$, there are two consequences: 1) the agent does not receive the resource, and 2) the agent pays a penalty (fine) $c \geq 0$.
Let $\alpha$ denote the allocation policy with $\alpha_i(f,\pazocal{A}', \phi)=1$ if agent $i$ receives the resource, and 0 otherwise.
Further, let $\caughtLying_i = 1$ if
agent $i$ is audited and $\z_i' \ne \z_i$ (the agent is caught lying) and 0 otherwise; note that since the audit policy is stochastic, $\caughtLying_i$ is a random variable.
We assume that an agent obtains a value of $1$ for receiving the resource and $0$ otherwise.
Consequently, the agent's utility is $u_i(\pazocal{A}') = \alpha_i(f,\pazocal{A}',\phi)(1-\caughtLying_i) - c \caughtLying_i$.

This game between a principal and agents can be expressed as the following sequence of events:
\begin{enumerate}
    \item The principal knows $D$, $n$, $\inter$, $\alpha$, $c$, and $f$, and announces an audit policy $\phi$.
    \item Realizations of $n$ agents are drawn i.i.d.\ from $D$. Each agent knows its own type $(\x_i, \z_i)$, $D$, $n$, $\inter$, $\alpha$, $c$, $\phi$, and $f$, but does not know the types of other agents. 
    \item Agents simultaneously submit their reported type $(\x_i, \z_i')$, where $\z_i'$ need not equal $\z_i$. 
    \item The principal audits up to $B$ agents, according to $\phi$. Any agent $i$ found to have reported $\z_i' \neq \z_i$ is removed from consideration (not allocated the resource), and pays a fine of $c$.
    \item The remaining agents are distributed a resource according to $\alpha$.
\end{enumerate}
Note that if an agent $i$ is found to be dishonest through an audit in the top-$k$ allocation setting, \emph{another agent would receive the resource in place of $i$}.

The goal of the principal is to achieve truthful reporting of types by the agents in an (approximate) Bayes-Nash equilibrium, or (approximate) \emph{Bayes-Nash incentive compatibility (BNIC)}. Formally:
\begin{definition}($\varepsilon$-BNIC)
An audit policy $\phi$ is $\varepsilon$-Bayes-Nash incentive compatible ($\varepsilon$-BNIC) if for all $i$ and $\aj_i$,
\begin{align*}
&\mathbb{E}_{\pazocal{A}_{-i} \sim D}[u_i(\aj_i, \pazocal{A}_{-i})|f,\phi,\alpha] \\
&\quad \ge \mathbb{E}_{\pazocal{A}_{-i} \sim D}[u_i(\aj_i',\pazocal{A}_{-i})|f,\phi,\alpha] - \varepsilon \quad \forall \aj_i': \x_j' = \x_j.
\end{align*}
$\phi$ is BNIC if it is $0$-BNIC.
\end{definition}

We consider two problems in this setting.
First, since it is in general impossible to induce BNIC, as we show below, we aim to identify an \emph{optimal} audit policy, defined as follows.
\begin{definition}\label{def:opt}(Optimal)
An audit policy $\phi$ is optimal if $\phi$ induces an $\varepsilon^{*}$-BNIC, and there does not exist another policy $\phi'$ for which truthful reporting is an $\varepsilon$-BNIC with  $\varepsilon < \varepsilon^{*}$.
\end{definition}
In other words, the optimal $\phi$ induces the least incentive to lie among all policies.\footnote{%
To avoid confusion, note that the principal could have other objectives, and our definition of optimality is specific to inducing the ``best'' approximation of BNIC.}
As a consequence, if we find that an optimal policy is not BNIC, then no policy can be.
Our second problem is to determine the smallest $\epsilon$ that can be induced by an audit policy. We show that in general, these problems have differing complexity.

Before proceeding with a general analysis, we make three observations about our model:
    1) if $B = n$, any score function $f$ can be made BNIC; 
    2) if $k \in \{0, n\}$, the top-$k$ case is trivially BNIC; and
    3) if $(1 + c)(B/k) \geq 1$, the top-$k$ case is again trivially BNIC.

%% file: NeedForAudit.tex





We begin by showing that without auditing the self-reported features (equivalently, when the audit budget $B=0$), ensuring BNIC amounts to ignoring $\z$ altogether whenever we use a deterministic scoring function $f$.
Since self-reported features may be important in determining priority of individuals for resources, this impossibility motivates a careful treatment of optimal auditing, which follows.

\begin{proposition}
\label{thm:impossibility}
     Suppose $B=0$.
     Then, both the top-$k$ and threshold mechanism are incentive compatible iff $f(\x, \z) = f(\x)$.
     Moreover, in the threshold setting, BNIC can be achieved only if $c>0$.
\end{proposition}
Due to space constraints, this and other full proofs are deferred to the supplement.

%% file: Optimal_Audit_Policy.tex
\section{Design of Optimal Audit Policies}

The problem of incentivizing truthfulness via auditing can be broken into two primary components: design and verification. The first component, design, is the construction of \emph{optimal} or approximately \emph{optimal} audit policies.
The second, verification, focuses on computing the maximum incentive to lie under an \emph{optimal} audit policy, denoted as $\varepsilon^*$.
Although both problems are in general hard, we show that verification is intrinsically ``harder" in the sense that in a wide range of settings optimally auditing agents is tractable, but computing $\varepsilon^*$ remains hard.
The focus of this section is on design. In particular, we exhibit a simple audit policy which is guaranteed to be optimal under the threshold allocation setting, and approximately optimal under the top-$k$ allocation setting. 

We begin with some remarks and notation that will be subsequently used in characterizing the optimal audit policies.
When selecting which agents to audit, the principal is unaware of each agent's true type $\aj_i = (\x_i, \z_i)$, and sees only the reported type $\aj_i' = (\x_i,\z_i')$. 
Since the principal is interested in minimizing the marginal gain that \emph{any} agent can achieve from lying, agents' true types must be considered through the lens of worst-case analysis. 
Note that the type with the largest incentive to report $(\x_i, \z_i')$ is the type with the lowest scoring $\z$, given \emph{known} type $\x_i$ (denoted as $\aj_i^* = (\x, \z_i^*)$. 
From the principal's perspective, any agent reporting $(\x_i, \z_i')$ must be assumed to have true type $(\x_i, \z_i^*)$.

With this in mind, agent reports can be classified as one of the following: a sure-truth, a sure-lie, or suspicious.
Sure-truths are reports which are guaranteed to be honest (e.g. $\z_i' = \z_i^*$).
Sure-lies are reports which are guaranteed to be false (these are only of the form $h(\x_i, \z_i') = 0$).
Suspicious reports are those with an unknown truth value. 
The following two definitions formalize these observations.

\begin{definition}\label{def:min_type}(Minimum Type)
    For any known partial type $\x_i$, we say the \emph{minimum type} of $\x_i$ is ${\aj_i^* = (\x_i, \z_i^*) = \arg\min_{\z\in I^s: h(\x_i, \z) > 0}f(\x_i, \z)}$.
\end{definition}
\begin{definition}\label{def:sus}(Suspicious) 
    We say a type $\aj_i'$ is \emph{suspicious} if the \emph{minimum type} $\aj_i^*$ has a strictly lower chance of being allocated a resource barring auditing, i.e.,
    ${\mathbb{E}_{\A_{-i}}\big[\alpha_i\big(f, \A_{-i}\cup\{ \aj_i'\}\big)\big] > \mathbb{E}_{\A_{-i}}\big[\alpha_i\big(f, \A_{-i}\cup\{ \aj_i^*\}\big)\big]}$
\end{definition}

The key point here is that the principle should never waste an audit on a sure-truth, and when looking at incentive compatibility (i.e. single deviations from collective truth-telling), there is at most one sure-lie in any set of reports, which should be audited with probability 1.
The more interesting question regarding audit polices is; what to do with \emph{suspicious} reports.

\subsection{Threshold Allocation}
Recall that in the threshold allocation setting, an agent receives a resource if $f(\x,\z') \ge \theta$, where $(\x,\z')$ is the agent's reported type.
We first show that, in general, optimal auditing under threshold allocation is NP-hard in general, but is tractable if and only if identifying sure-truths is tractable.
The hardness of auditing stems from the possibly arbitrary relationship between the distribution $D$ and the score function $f$.
\begin{theorem}\label{thm:thresh_audit_hard}
    For a given set of $n$ reports $\A'$ and a budget $B$, computing an optimal audit policy is NP-hard.
\end{theorem}
\begin{proof}[Proof Sketch]
    This result stems from the observation that the principal would never want to ``waste'' an audit on an agent whose report is guaranteed to be truthful. 
    For example, suppose agent $\aj_1 = (\x_1, \z_1)$ reports type $\aj_1' = (\x_1, \z_1')$ with $f(\x_1, \z_1') \geq \theta$. Suppose further that for all $\z$ with $f(\x_1, \z) < \theta$, we have ${h(\x_1, \z) = 0}$. 
    Then the principal is certain that agent $1$ is truthful since this agent's true type could not have scored below the threshold. 
    Due to this dependency on the underlying distribution, one can encode a SAT formula into the distribution such that determining if there exists a $\z$ such that $h(\x_1, \z) > 0$ and  $f(\x_1, \z) < \theta$ is equivalent to determining the satisfiability of the SAT instance.
\end{proof}

To better understand the nature of the problem of characterizing an optimal audit policy, consider the following simple example.
\begin{example}\label{ex:1}\normalfont
Suppose there are two agents with one \emph{known} and one \emph{self-reported} binary feature, and suppose that $z = 1$ if $x=1$, and can be either 0 or 1 according to some prior distribution if $x=0$.
Further, suppose that $f(x,z) = z$ and $\theta = 1/2$, which means that an agent receives the resource iff $z=1$.
Now, suppose that $B=1$ and the principal observes two types: $(1,1)$ and $(0,1)$.
Clearly, the principal would not audit the former, since $x=1$ already implies that the agent is honest, but would audit the latter.
This simple example suggests that one could expect an optimal audit policy to depend in rather complex ways on the observed types $\pazocal{A}'$.
\end{example}

However, we show that a simple policy of uniformly auditing all \emph{suspicious} agents (Definition \ref{def:sus}), is optimal. 
We call this policy \unf, and define it formally next.
\begin{definition}
    (\unf) For a given set of reports $\A'$, let $G(\A')$ be the set of all agent's whose reports are \emph{suspicious}.
    Given budget $B$, the \unf~ audit policy audits each $\aj' \in \A'$ with probability
    \begin{align*}
        \phi_{i}(\A') = 
        \begin{cases} 
          1 & \text{ if } h(\aj_i') = 0\\
          \min\big(\frac{B}{|G(\pazocal{A}')|}, 1 \big) & \text{ if } \aj_i' \in G(\pazocal{A}'),\\
          0                                            & \text{otherwise}
       \end{cases}
    \end{align*}
\end{definition}
Next, we show that in the threshold allocation setting, this \unf\ audit policy is optimal.

The intuition for the optimality of \unf\ comes from the fact that any type $\z$ can report any other type $\z'$ at no cost.  This means that any lie that gets an agent above the threshold is equivalent, modulo auditing. Thus, if an audit is non-uniform (as long as the reported type is above the threshold), some lies become more valuable than others, and we should shift auditing to those lies (more precisely, to agents who feature such lies). 
The discontinuity arises by observing a sure-lie (i.e., $h(x, z) = 0$); only in this case do we know which agent was dishonest, and can thus place higher audit weight on this agent without increasing the value of lying for any other agent.

Note that this implies optimal auditing is equivalent to identifying sure-truths.

\begin{theorem}\label{thm:bayes_nash_best_policy}
    In the threshold allocation setting, for any score function $f$, \unf~is an \emph{optimal} audit policy.
\end{theorem}

\begin{proof}[Proof Sketch]
    For the sake of illustration, we demonstrate how this result holds in the cases of a discrete distribution over agent types. An identical idea holds for continuous features, although the technical details differ.
    
    
    When analyzing \eBNIC, we are considering the value that any agent gains when deviating from a truthful reporting, while all other agents remain truthful, i.e. we consider this case when at most one report is dishonest. In any set of reports $\A'$, if the principal sees a sure-lie, they are immediately aware of the dishonest agent's identity and should exclusively audit that agent, since all other agents are guaranteed to be truthful.
    
    The principal's objective is to minimize the expected gain of any type $\aj_i$ misreporting their type as $\aj_i'$, when all other agents are truthful. Note that when all agents, aside from agent $i$ are honest, the set of reported types $\A' = \A_{-i} \cup \{ \aj_i'\}$ (where $\A_{-i}$ is the set true types for all other agents). As such, we can express the minimum expected gain of misreporting, achievable by any audit policy $\phi$, as
    \begin{align}
        &\varepsilon = \min\limits_{\phi}~\max\limits_{\aj_i', \aj_i}~\bigg(\underset{\A_{-i}}{\mathbb{E}}\big[\alpha_i(f, \A') - \alpha_i(f, \A)\big]\\
        &\quad \quad\quad\quad\quad\quad - \underset{\A_{-i}}{\mathbb{E}}\big[\big(\alpha_i(f, \A') + c\big) \phi_i(\A')\big]\bigg)
    \end{align}
    Where term $(1)$ the expected difference in the allocation decision between agent $i$ falsely reporting $\aj_i'$ or truthfully reporting $\aj_i$, and term $(2)$ represents the expected cost of being caught lying when reporting  $\aj_i'$. 
    Making use of two simple observations, we can simplify this equation. First, in the threshold setting, agents know both their own type and the threshold $\theta$, thus agent $i$ knows the allocation decision on both the true type $\aj_i$, and reported type $\aj_i'$, meaning that the expectations on $\alpha$ can be dropped. Second, we need only consider this term for \emph{suspicious} agents, so we may assume that $\alpha_i(f, \A') = 1$ and $\alpha_i(f, \A) = 0$. With this, the equation can be simplified to
    \begin{align*}
       \varepsilon = \min\limits_{\phi}~\max\limits_{\underset{f(\aj_i') \geq \theta > f(\aj_i)}{\aj_i', \aj_i}}~1  - (1 + c)\mathbb{E}_{\A_{-i}}\big[\phi_i(\A') \big]
    \end{align*}
    Thus $\varepsilon$ is solely determined by the value of $\mathbb{E}_{\A_{-i}}\big[\phi_i(\A') \big]$ for any \emph{suspicious} type $\aj_i'$. Let 
    \begin{align*}
        G(\A') = \{ &(\x, \z') \in \A': f(\x, \z') \geq \theta \text{ and } \exists \z^* \text {s.t. } f(\x, \z^*) < \theta \\
    &\text{ and } h\big((\x, \z^*) \big) > 0 \text{ and } h\big((\x, \z')\big) > 0\}
    \end{align*} 
    be the set of \emph{suspicious} types in $\A'$. In the case when agent features are distributed according to a discrete distribution, this expectation can be expressed as
    \begin{align*}
        \mathbb{E}_{\A_{-i}}\big[\phi_i(\A') \big] =& \sum_{\A_{-i}}\phi_i(\A') Q(\A_{-i}) \\
        =& \sum_{\A_{-i}}\min\bigg(1, \frac{B}{G(\A')} \bigg) Q(\A_{-i}),
    \end{align*}
    where $Q(\A_{-i})$ is the probability of any realization of the specified types of agents other than $i$ induced by $D$.
    The probability which sure-lies are audited has no effect on the value of other lies, and thus sure-liescan be audited with probability $1$. Moreover, the sum is equal for any two \emph{suspicious} agents with $h(\aj') > 0$. In each set of reports $\A'$, the principle fully spends their budget (or audits all \emph{suspicious} types with probability $1$) and $Q(\A_{-i})$ is independent of the type agent $i$ reports. 
    Thus, for any policy different from \unf, at least one audit weight must be changed, i.e., $\phi_i(\A') \neq \min\big(1, \nicefrac{B}{G(A')})$, for some $i$ and some $\A'$. As a result of the tightness and independence of $Q(\A_{-i})$, this change of audit weight could only result in a (not necessarily strict) increase in the expected gain of misreporting for any agent type. 
\end{proof}

Note that while optimal, \unf\ is in general intractable because of the combinatorial structure of such policies that may be induced by $h(\cdot)$.
However, we now show that for sufficiently well-behaved $h$ and $f$ we can compute \unf\ efficiently.
\begin{theorem}\label{thm: unf_is_tract}
   The audit policy \unf\ can be computed in polynomial time if for any report $(\x, \z')$, it can be efficiently determined if $(\x, \z')$ is a sure-truth, (i.e. there exists a self reported type $\z^*$, such that $h(\x, \z^*) >0$ and $f(\x, \z^*) < \theta$).
\end{theorem}


\subsection{Top-$k$ Allocation}

We now turn our attention to selecting the \emph{optimal} audit policy when resources are given to the $k$ highest scoring agents.
In this case, the optimal policy no longer admits a clean characterization.
The main challenge is that now there are far more complex interdependencies among agents' benefits from lying, other agents' reports, and the audit policy.
For example, if an agent in the top-$k$ is caught lying, another agent would now receive the resource.
Instead, we study a natural adaptation of \unf\ to this setting, and exhibit an additive approximation bound for its optimality.
We then show that if we use dominant strategy incentive compatibility (defined formally below) as a solution concept in place of BNIC, uniform auditing is  optimal even in this setting.


We begin by showing that optimal auditing in the top-$k$ setting is NP-hard even when sure-truths are identifiable in constant time.
\begin{theorem}\label{thm:hard_audit_top_k}
    In the top-$k$ allocation setting determining which agents should be audited is NP-hard,
    even for $n = 4$ agents, monotone $f$, uniform $D$, and even if sure-truth can be identified in constant time. 
\end{theorem}
\begin{proof}[Proof Sketch]
We can encode an instance of Vertex Cover into $f$ such that agents with a self-reported type, which constitutes a vertex cover, ranks in the top-$k$ with extremely low probability, while all other types have score proportional to number of vertices that their self-reported type ``covers".
For a sufficiently small budget and penalty for lying, there will be agents whose expected value of lying (even if never audited) is smaller than agents who receive the highest probability weight. As such, the principal must determine which agents should receive zero audit weight, which is NP hard due to the encoding of VC. 
\end{proof}

Now, consider a variant of the \unf\  policy in the top-$k$ setting where we uniformly at random audit agents who have scores in the top $k$.
We first define this policy formally.

\begin{definition}
    (\unfk) For any set of reported types $\A'$, let \unfk\ denote the policy of auditing each of the top-$k$ agents $($refereed to as the set $T_k\subset \A')$ with probability $\min(1, \nicefrac{B}{k})$.
\end{definition}
Next, we show that \unfk\ admits an additive approximation of an optimal audit policy in the top-$k$ setting.  Recall that multiplicative approximations are in general NP-hard to achieve.

\begin{theorem}
    Let $\phi$ denote the audit policy \unfk. Then the maximum utility gained by lying under $\phi$ is no more than $\max\big(0, 1 - \frac{(1 + c) B}{k}\big)$ greater than that of the optimal audit policy, and this bound is tight.
\end{theorem}
\begin{proof}
    This is the result of simple worst case analysis on the  expected value of lying, which can be expressed as
   \begin{align*}
        &\mathbb{E}_{\A_{-i}}\big[\alpha_i(\A')\big(1 - \phi_i(\A')\big) - c\phi_i(\A') - \alpha_i(\A)\big] \\
        =& \mathbb{P}(\aj_i' \in T_k)\mathbb{E}[\phi_i(\A') | \aj_i'\in T_k] -c\mathbb{E}[\phi_i(\A')] + \mathbb{P}(\aj_i\in T_k)
   \end{align*}
   In the worst case, the expected value of lying could be $0$ for all agents. However, the uniform audit policy will have expected value of lying equal to \( \mathbb{P}(\aj_i'\in T_k)(1 - (1 + c) \frac{B}{k})  - \mathbb{P}(\aj_i\in T_k)\). Which again in the worst case is equal to ${((1 - (1 + c)\frac{B}{k})}$
   
   This bound is tight to within any small $\beta > 0$. To see this, construct an instance with 3 agents of types $x\in\{0, 1\}, z\in\{0, 1\}$. Where $f(x, z) = x\land z$. Let $\mathbb{P}(x = 1, z = 1) = \beta$ and the rest have probability $\frac{1 - \beta}{3}$. Let $B = 1$ and $k = 2$. Then an optimal audit policy yields $\varepsilon^* = 0$, but uniformly auditing the top-$k$ yields $\varepsilon =((1 - \beta^2)(1 - (1 + c)\frac{B}{k})$.
\end{proof}

A major part of what makes auditing difficult is the dependence on the distribution.
We now consider an alternative solution concept which eliminates this dependence:
$\varepsilon$-Dominant Strategy Incentive Compatibility ($\varepsilon$-DSIC). 
Specifically, under $\varepsilon$-DSIC the principal aims to design a policy under which truthful reporting is (approximately) optimal for agents regardless of other agents' types. 
\begin{definition}
An audit policy $\phi$ is $\varepsilon$-DSIC if for all $i$ and $\aj_i$,
\begin{align*}
&\mathbb{E}[u_i(\aj_i, \pazocal{A}_{-i})|f,\phi,\alpha, \A'_{-i}] \\
&\ge \mathbb{E}[u_i(\aj_i',\pazocal{A}_{-i})|f,\phi,\alpha, \A'_{-i}] - \varepsilon \ \forall \aj_i': \x_j' = \x_j \text{ and } \forall \A'_{-i}.
\end{align*}
\end{definition}


\begin{theorem}\label{thm:unfk_opt_for_DNIC}
    In the top-$k$ setting, \unfk\ yields $\varepsilon^*$-DSIC with an optimal $\varepsilon^*$.
\end{theorem}

\begin{proof}[Proof Sketch]
    
    In the top-$k$ setting the key difference from \eBNIC\ is that for any realization $\A_{-i}$ and any set of corresponding reports $\A_{-i}'$, agent $i$ knows the allocation decision on both their true type $\aj_i$ and any reported type $\aj_i'$.
    This certainty of outcomes it precisely what made all \emph{suspicious} reports equivalent in the threshold case.
    Using a similar argument for the optimality of \unf\ in the threshold case, we can see that \unfk\ is optimal in the top-$k$ case.
\end{proof}

%% file: Hardness.tex
\section{Verification of Policy Effectiveness}

In the previous section we showed that in many circumstances we can fully characterize the optimal audit policy, and it can be efficiently computed for a broad range of settings.
We now consider the problem of \emph{verification}, that is, computing the smallest $\epsilon^*$ that we can achieve for an optimal audit policy.
We show that this problem is hard even when auditing is easy.
Subsequently, we first show that we can often effectively approximate this problem, and then exhibit special cases in which we can even compute this $\epsilon^*$ efficiently.


\subsection{Complexity of Verification}

In the threshold setting, we will show that computing the minimum $\varepsilon^*$ inducible by any policy is \#P-hard, even in cases when optimal auditing is tractable.
This complexity stems from \emph{both} the score function $f$ and distribution $D$. Intuitively, these uniquely define both the set of agent types which are considered \emph{suspicious} and the probability that a \emph{suspicious} type will occur. As \emph{suspicious} types are more likely to occur, the probability that any particular agent is audited decreases. Thus, we can encode ``hard" problems into $f$ or $D$ where agent types (binary vectors) correspond to satisfying assignments of the encoded problem.
We can also observe that if the number of possible agent types is polynomial, then the problem is trivially tractable through brute force search.

We show here hardness in terms of $f$; a similar construction works to show the hardness in terms of $D$. 
In this construction, optimal auditing is easy even in the top-$k$ case.
\begin{theorem}\label{thm:thresh_hard}
    In both the threshold and top-$k$ setting, computing the minimum $\varepsilon$ inducible by any audit policy is \#P-hard, for both continuous and discrete features, even when the feature distribution is uniform, there are only $2$ agents, and $f$ is both monotone and binary.
\end{theorem}
\begin{proof}[Proof Sketch]
For this proof sketch we will work in the setting of threshold allocation and discrete features, similar logic holds in the other cases. We reduce from \#VC.
For a graph ${G= (V, E)}$, let $D$ be uniform and agents be ${\aj = \langle x_1, ..., x_{|V|}, z_1\rangle}$, for $x, z\in \{0, 1\}$. Let $\theta = \frac{1}{2}$ and set 
\begin{align*}
    f(\x, z) = \big(\bigwedge_{(v_r, v_t) \in E}(x_r \lor x_t)\big)\land z_1.
\end{align*}
Under this construction of $f$ we see that an agent scores $f(\aj) = 1$ if and only if $\x$ constitutes a vertex cover and $z = 1$. Thus when $B = 1$ and $n=2$, if agent $1$ scores below $\frac{1}{2}$ and is considering misreporting their type, they are audited with lower probability if $f(\aj_2) = 1$. Since $D$ is uniform, the probability of this occurring is equivalent to the number of vertex covers of $G$.
\end{proof}

In addition to hardness of checking BNIC, we can show that it is even hard to multiplicatively approximate an $\varepsilon$-BNIC in the threshold and top-$k$ settings.

\begin{theorem}\label{thm:approx_hard}
    Multiplicatively approximating to any constant factor the smallest $\varepsilon$ such that there is an $\varepsilon$-BNIC audit policy, in both threshold and top-$k$ allocation is NP-hard even for $\Theta(1)$ agents.
\end{theorem}

\begin{proof}
    This result is a straightforward consequence of the construction in the proof of Theorem \ref{thm:thresh_hard}. In that proof we encode an NP-hard problem into an instance of our problem, and show that determining if truthful reporting is BNIC is equivalent to counting the number of satisfying assignments of vertex covers.
    If we reduce instead from an Unambiguous-SAT instance ($f$ is no longer monotone), then the mechanism is BNIC if and only if the formula has exactly one satisfying assignment. 
    This would imply that $\varepsilon = 0$ if and only if the U-SAT instance has no satisfying assignment, and any multiplicative factor $\varepsilon$ would likewise be zero, immediately indicating the satisfiability of the U-SAT instance.
\end{proof}

Note that \unfk\ is the optimal audit policy in these cases, implying that not only is verification of an optimal policy hard, but also verification of \unfk\ is also in general hard.

In summary, the problem of \emph{checking} whether a particular setting is \eBNIC\ is hard, even in instances when auditing is tractable. 
To further outline the relation of the complexity of both problems we make the following observation.
\begin{theorem}\label{thm: ver_p_implies_audit_p}
    In the threshold allocation setting, verification being in P implies optimal auditing is also in P.
\end{theorem}
Next, we turn to positive results.
To begin, we now show that when agents' \emph{minimum} type can be efficiently computed we can achieve a probabilistic bound on the value of lying in polynomial time, via Monte Carlo simulations.
\begin{theorem}\label{thm:thresh_approx_sample}
    Suppose that $\varepsilon^*$ is the minimum value for which \unf~is $\varepsilon^*$-BNIC. Then, for any $\gamma \in \Theta(1)$, performing $n^{\gamma}$ rounds of Monte-Carlo sampling will yield a value $\varepsilon'$, such that $\varepsilon'  = \varepsilon^* \pm \Theta\big(\nicefrac{1}{\sqrt{n^{\gamma} - 3}}\big)$ with probability at least $1 - \nicefrac{1}{n^2}$. This can be done in time $\Theta(n^{\gamma + 1})$.
\end{theorem}
Observe from Theorem~\ref{thm:thresh_approx_sample} that as $n$ increases, the error of approximation tends towards 0 with probability tending towards 1.
Next, we consider special cases in which verification is tractable.

%% file: Special_Cases.tex
\subsection{Tractable Special Cases}
Thus far, our results are negative when it comes to checking incentive compatibility, and mixed in terms of devising an optimal audit policy.
We now proceed to identify a number of special cases in which we can check incentive compatibility in polynomial time.
In the threshold setting, we focus on checking \eBNIC\ for a \unf\ audit policy, which we showed earlier is optimal, while in the top-$k$ setting we focus on the \unfk\ audit policy.
We consider, in particular, three common machine learning models for $f$: linear, piecewise linear, and logistic (sigmoid) functions. 
Throughout, we assume that distributions over types are sufficiently well behaved, in that it is tractable to compute probabilities of intervals.

We begin by showing that verification is tractable in any instance in which the CDF (CMF) of $h$ can be computed over the set of \emph{suspicious} agent types. As can be surmised from the complexity results regarding verification, the ``hardness" of the problem stems from determining the probability that an agent's true type is suspicious. However, when this can be computed efficiently, so can $\varepsilon^*$.
\begin{theorem} \label{thm: tract}
    Let $U = \{(\x, \z')\in\inter : f(\x,\z') \geq \theta, h(\x, \z) \neq 0, \text{ and } \exists (\x, \z^*) \text{ with } f(\x, \z^*) < \theta \text{ and } h(\x, \z^*) \neq 0 \}$. If $\mathbb{P}_{\aj\sim D}\big(\aj\in U \big)$ can be efficiently computed, then so can $\varepsilon^*$.
\end{theorem}

\begin{proof}[Proof Sketch]
    Let $p_U = \mathbb{P}_{\aj\sim D}(\aj \in U)$. Suppose an agent initially scores below the threshold, then this agent's only means for allocation is to report a type in $U$. Moreover, \unf\ only audits agents in $U$ and does so uniformly. Thus, for a given realization, the more agents with true types in $U$, the lower the probability that the dishonest agent is audited.  More specifically, suppose that some agent $\aj_i = (\x_i, \z_i)$, with $f(\aj_i) < \theta$, is able to falsely submit $\aj_i' = (\x_i, \z_i')$ with $f(\aj_i') \geq \theta$. Then, this agent's expected marginal gain is,
    \begin{align*}
        &\mathbb{E}_{\A_{-i}}[u_i(\aj_i, \aj_i') | f, \alpha, \phi] = \mathbb{E}_{\A_{-i}}[1 - (1+c)\phi_i(\A')]\\
        &= 1 - (1+c)\sum_{\ell = 0}^{n-1}\binom{n-1}{\ell}p_U^{\ell}( 1- p_U)^{n - \ell - 1}\min\big(1, \nicefrac{B}{\ell + 1})
    \end{align*}
    Since, under \unf\, all dishonest reports have either value $0$ or value $\mathbb{E}_{\A_{-i}}[u_i(\aj_i, \aj_i') | f, \alpha, \phi]$, we need only compute this single sum, for any agent type, and have found $\varepsilon$. Moreover, \unf is optimal and thus $\varepsilon = \varepsilon^*$.
\end{proof}

In both the discrete and continuous case, when $\mathbb{P}(\aj \in U)$ can be computed exactly, verification is tractable. 
Next, we give a sufficient condition on this, and present several tractable special cases.
\begin{definition}
    We say a PDF $h$ is \emph{well-behaved} if $h$ is zero on a polynomial number of $s + d-$dimensional maximal intervals, and over any any interval $[a, b]\subset \R$, the value of $\int_a^b h(\x, \z) d z_r$ and $\int_a^b h(\x, \z) d x_t$ for observed features $r$ and unobserved features $s$ have closed-form solutions derivable in polynomial-time w.r.t. $(n, B, s, d, \log(c))$.
\end{definition}

\begin{remark}
    Many commonly used distributions, such as uniform and exponential, are \emph{well-behaved}. In many other common cases, such as Gaussian, we can obtain a good numerical approximation, so that the approaches below can approximately apply in these also. We formalize this below.
\end{remark}

As we show next, in the threshold case, checking $\varepsilon$-BNIC is easy for piecewise linear and logistic score functions as long as the distribution over types is \emph{well-behaved}. 
For top-$k$, we need a much stronger assumption that types are distributed uniformly to obtain comparable positive results.
Each proof proceeds as follows (see the supplement for details). The score function $f$ partitions $\inter$ into two disjoint regions, one of which is $U$ (the set of suspicious types). We then show that one of the two partitions it is easy to compute the CDF, as long as $h$ is \emph{well-behaved}. Once this is done, we have computed either $p_U$ or $1 - p_U$ and from here Theorem \ref{thm: tract} directly implies that $\varepsilon^*$ is computable in polynomial time.


\begin{definition}
    A function $f:\R^n \rightarrow \R$ is said to be \emph{Piecewise Linear} if for some partition of $\R^{d + s}$ into disjoint rectangular regions, given by $P = \{ L_1, ..., L_m\}$ the function $f\big|_{L_s}:\R^{d + s} \rightarrow \R$ is linear for each $L_s\in P$.
\end{definition}

\begin{corollary}
\label{cor:piecewise-nashEq}
        Suppose the distribution of agent types is well-behaved and $f$ is piecewise linear, or logistic, and $\alpha(f, \A')$ is threshold allocation.
        Then determining the minimum $\varepsilon \geq 0$ such that \unf~is $\varepsilon$-BNIC, can be done in polynomial time.
\end{corollary}

\begin{corollary}\label{thm:unf_on_PWL_top_k}
        Suppose the distribution of agent types is uniform. Suppose further that $f$ is piecewise-linear, or logistic, and $\alpha(f, \A')$ is top-$k$ allocation. Then determining the minimum $\varepsilon \geq 0$ such that \unfk~is $\varepsilon$-BNIC, can be done in polynomial time.
\end{corollary}

For many common continuous distributions, such as Gaussian, only a numerical approximation of $p_U = \mathbb{P}(\aj \in U)$ can be computed. 
Our final result is to quantify the error in $\varepsilon^*$, in terms of the additive numerical error $\gamma$ in $p_U$.
\begin{theorem}\label{thm:approx_error_tractable_cases}
        Suppose with error $\gamma$ we have a numerical approximation $p_U' = p_U \pm \gamma$. Then we can compute $\varepsilon' = \varepsilon^* \pm (n - B)\binom{n - 1}{B - 1}\int_{p_U}^{p_U + \gamma}x^{B - 1}(1 - x)^{n - B} dx$.
\end{theorem}
Although the error term looks messy, it is tight and in general small relative to $\gamma$, which itself is also in general a small value. As an illustration, when we have error $\gamma = 4.44\text{E}^{-4}$, a typical absolute error for a standard Gaussian, and $n = 1000, B = 250$, and  $p_U = 0.6$, then $\epsilon' = \epsilon^* \pm 6\text{E}^{-60}$.

%% file: conclusion.tex
\section{Conclusion}

We study the problem of auditing self-reported attributes in resource allocation settings from two perspectives: 1) the complexity of checking whether a particular audit policy is incentive compatible, and 2) characterizing and computing an audit policy that minimizes incentives to lie.
We find that checking incentive compatibility is, in general, hard. However, in settings where resources are assigned by thresholding the individual's computed score, a uniform audit policy, particularly appealing for its simplicity, is optimal. 
In addition, we show that in two important classes of score functions, piecewise linear and logistic, we can check incentive compatibility in polynomial time under some assumptions on the distribution of agent types. 

A number of open questions remain.
While we show that computing an optimal audit policy in the setting where resources are allocated to the top-$k$ scoring agents is hard, it may be possible to achieve a better approximation of optimal than what we exhibit for the uniform policy.
Moreover, our model presumes that agents incur no direct costs of misreported preferences besides the endogenous costs of being audited.
In practice, there may be both cognitive and tangible costs involved, and these can be considered as an extension to our model.
Finally, we assume that the distribution over agent types is known a priori, whereas it likely needs to be learned from data.